\documentclass[twocolumns, letterpaper]{aa}
\usepackage{graphicx}
\usepackage{txfonts}
\begin{document}
   \title{X-ray flares from the ultra-luminous X-ray source in NGC\,5408}

%   \subtitle{I. Overviewing the $\kappa$-mechanism}

%\author{Motch, Read, Soria, Stevens---order to be determined}
   \author{R. Soria
          \inst{1,2},
		C. Motch\inst{2},
			A. M. Read\inst{3}
          \and
          I. R. Stevens\inst{4}}
%\fnmsep\thanks{Just to show the usage
%          of the elements in the author field}
%          }

   \offprints{R. Soria}

   \institute{Mullard Space Science Laboratory, University College London, 
	Holmbury St Mary, Surrey, RH5 6NT, UK\\
              \email{Roberto.Soria@mssl.ucl.ac.uk}
         \and
             Observatoire Astronomique, UMR 7550 CNRS, 
11 rue de l'Universit\'{e}, 67000 Strasbourg, France
%             \email{motch@astro.u-strasbg.fr}
         \and
Department of Physics and Astronomy, Leicester University, Leicester LE1 7RH, UK
\and
School of Physics and Astronomy, University of Birmingham, 
Edgbaston, Birmingham B15 2TT, UK
%             \email{motch@astro.u-strasbg.fr}
%             \thanks{The university of heaven temporarily does not
%                     accept e-mails}
             }

   \date{Received April 23, 2004 }

   \abstract{
We have studied an ultra-luminous X-ray source (ULX) 
in the dwarf galaxy NGC\,5408 with 
a series of {\it XMM-Newton} observations, between 2001 July 
and 2003 January. We find that its X-ray spectrum is best fitted with  
a power law of photon index $\Gamma \approx 2.6$--$2.9$ 
and a thermal component with blackbody temperature 
$kT_{\rm bb} \approx 0.12$--$0.14$ keV. These spectral features, 
and the inferred luminosity $\approx 10^{40}$ 
erg s$^{-1}$ in the $0.3$--$12$ keV band, 
are typical of bright ULXs in nearby dwarf galaxies. 
The blackbody plus power-law model is a significantly better fit than 
either a simple power law or a broken power law (although the latter model 
is also acceptable at some epochs). Doppler-boosted emission 
from a relativistic jet is not required, although we cannot rule out 
this scenario. Our preliminary timing analysis shows flaring behaviour 
which we interpret as variability 
in the power-law component, on timescales of $\sim 10^2$ s. 
The hard component is suppressed during the dips, 
while the soft thermal component is consistent with 
being constant. 
The power density spectrum is flat at low frequencies, has a break 
at $\nu_{\rm b} \approx 2.5$ mHz, and has a slope $\approx -1$ 
at higher frequencies. A comparison with the power spectra of Cyg X-1  
and of a sample of other BH candidates and AGN suggests 
a mass of $\sim 10^2 M_{\odot}$. It is also possible that 
the BH is at the upper end of the stellar-mass class 
($M \sim 50 M_{\odot}$), in a phase of moderately super-Eddington 
accretion. The formation of such a massive BH via 
normal stellar evolution may have been favoured 
by the very metal-poor environment of NGC\,5408.

   }

\titlerunning{Flares from the ULX in NGC\,5408}

   \maketitle
%
%________________________________________________________________

\section{Introduction }

%%%%%%%%%%%%%%%%
Ultra-luminous X-ray sources (ULXs) are point-like sources, 
not including galactic nuclei and young supernova remnants, 
with apparent luminositites higher than the Eddington limit for 
a stellar-mass accreting black hole (BH), ie, with $L_{X}\ga 3 \times 
10^{39}$\,erg s$^{-1}$ (Fabbiano 1992; Colbert \& Mushotzky 1999; 
Roberts \& Warwick 2000). The true nature
of these objects is still open to debate. 
The fundamental issue is whether the emission is isotropic 
or beamed along our line-of-sight; 
in the latter case, the total luminosity would 
of course be lower than the isotropic value. 
A possible scenario for geometrical beaming 
is a super-Eddington mass inflow during phases of 
thermal-timescale mass transfer 
%in high-mass X-ray binaries, 
%and during transient outbursts in low-mass X-ray binaries 
(King 2002).
Relativistic beaming, associated with Doppler boosting, 
would instead be the effect of our looking into the jet 
of a microquasar (K\"{o}rding et al.~2002).
Alternatively, if the emission is isotropic and the Eddington limit 
is not violated, ULXs must be fuelled by accretion onto 
intermediate-mass BHs, with masses $\sim 10^{2}\,M_{\odot}$. 
Possible mechanisms of formation of such massive remnants include 
the collapse of massive population {\rm III} stars 
(Madau \& Rees 2001), or runaway stellar mergers in young clusters 
(Portegies Zwart et al.~2004).

A ULX in the starburst dwarf irregular galaxy NGC\,5408 
could help discriminate between these two alternatives.
Radio observations of this galaxy  
(Stevens et al.~2002) showed four main emission regions, 
mostly coincident with massive young star clusters 
and H$\alpha$ emission. 
%(Bohuski et al. 1972).
The X-ray emission is dominated by a single point source, 
with a luminosity $>10^{40}$ erg s$^{-1}$ 
(Fabian \& Ward 1993; Kaaret et al.~2003), located outside 
the H{\footnotesize II} regions, 
but apparently coincident with a weak, steep-spectrum 
radio source. On the basis of the X-ray, radio, and optical fluxes, 
Kaaret et al.~(2003) concluded that the most likely explanation 
for the ULX was beamed relativistic jet emission
from a stellar-mass BH. However, they could not
rule out an intermediate-mass BH model. 
In this paper we present the first results of our {\it XMM-Newton} 
study of the ULX in NGC\,5408, providing new insights 
into the nature of this source.

\section{Observations and data analysis}

NGC\,5408 was observed with the European Photon Imaging Camera (EPIC)
and the Reflection Grating Spectrometer (RGS) 
on board {\it XMM-Newton} in 2001 July--August, 2002 July and 2003 January 
(Table 1). The Optical Monitor was blocked because 
of a bright foreground star in the field.
In this paper, we present preliminary results 
from the EPIC data; more detailed results are left 
to further work.

The thin-filter, full-frame mode was used for the MOS 
and pn cameras. We processed the Observation Data Files 
with standard tasks in Version 5.4 of the Science Analysis 
System (SAS). After inspecting the background flux of each 
exposure, we rejected various intervals with background flares 
in Rev.~301 and Rev.~574.
%and (for the timing analysis only), 
%at the beginning of the MOS exposures in Rev.~305. 
A high, flaring background affected 
the whole exposures in Rev.~483: the average background 
count rate was $\approx 20\%$ of the ULX count rate 
in the source extraction region. We retained that exposure 
(the only one available for 2002), with the caveat that 
the timing analyses may be affected by larger errors. 
For all other exposures, 
the background rate in the selected good-time-intervals 
is typically $\sim 1$--$2\%$ of the source count rate.
Technical problems during Rev.~313 meant that the pn 
was not activated, while the usable part of the MOS exposures 
lasted only 6.9 ks despite a nominal exposure time of 7.5 ks.
We filtered the event files, selecting only the best-calibrated 
events (pattern $\leq 12$ for the MOSs, pattern $\leq 4$ for the pn), and 
rejecting flagged events. We also checked that the source was not piled-up,
with the SAS task {\tt epatplot}.

For each exposure, we used a 30\arcsec-radius source extraction 
region, with the background extracted from other regions 
in the same CCD. We built appropriate response functions with 
the SAS tasks {\tt rmfgen} and {\tt arfgen}. 
We fitted the background-subtracted 
spectra with standard models in {\footnotesize XSPEC v.11.3} (Arnaud 1996); 
owing to the uncertainties in the EPIC responses at low energies, 
we used only the $0.3$--$12$ keV range.
Firstly, we analysed the individual pn and MOS spectra from each observation. 
After ascertaining that they were consistent with each other, 
we coadded them with the method described in Page et al.~(2003), 
to increase the signal-to-noise ratio. 
Thus, we obtained one combined EPIC spectrum of the ULX 
from each epoch. With the same method, we also built an average 
EPIC spectrum of all the observations from 2001.
For the timing analysis presented here, 
we selected all the good-time-intervals when both the MOS and pn
were operating (except for the lightcurve in Rev.~313, 
for which only the MOS is available), and we retained the events 
in the $0.2$--$12$ keV range. We used standard tasks in 
{\footnotesize FTOOLS}
%\footnote{http://heasarc.gsfc.nasa.gov/ftools} 
(Blackburn 1995) to obtain background-subtracted lightcurves 
and power-density spectra.

%__________________________________________________ One column table
   \begin{table}
      \caption[]{{\it XMM-Newton}/EPIC observation log for NGC\,5408.}
         \label{table1}
\begin{centering}
         \begin{tabular}{l@{\hspace{0.3cm}}r@{\hspace{0.3cm}}r@{\hspace{0.3cm}}r@{\hspace{0.3cm}}r}
            \hline
            \hline
            \noalign{\smallskip}
            Date & Obs ID & \hspace{-0.2cm}Start time & \hspace{-0.2cm}Stop time & \hspace{-0.4cm}GTI\\[3pt]
%		&&&&(ks)\\[3pt]
            \noalign{\smallskip}
	    \hline
            \hline
            \noalign{\smallskip}
            \noalign{\smallskip}
		2001 Jul 31 & 0112290501 & pn: 12:49:45 & 14:13:05 & 3.8 ks\\
			& (Rev.~301) & MOS: 12:12:08 & 14:17:21 & 4.1 ks\\
            	2001 Aug 08 & 0112290601 & pn: 10:22:14 & 11:45:34 & 4.5 ks\\
			& (Rev.~305)&	  MOS: 09:43:01 & 11:48:13 & 7.4 ks\\
            	2001 Aug 24 & 0112290701 & pn: - & - & -\\
			&(Rev.~313)&	  MOS: 21:58:47 & 00:05:39 & 6.9 ks\\
            \noalign{\smallskip}
	    \hline
            \noalign{\smallskip}
		2002 Jul 29 & 0112291001 & pn: 08:23:12 & 10:08:22 & 4.9 ks \\
			& (Rev.~483)& MOS: 08:01:02 & 10:11:28 & 7.7 ks\\
            \noalign{\smallskip}
	    \hline
            \noalign{\smallskip}
		2003 Jan 28 & 0112291201 & pn: 00:06:06 & 01:29:57 & 3.0 ks\\
			& (Rev.~574)& MOS: 23:43:56 & 01:18:14 & 3.3 ks\\
            \noalign{\smallskip}
	    \hline
         \end{tabular}
\end{centering}
   \end{table}

\section{Main results}

\subsection{Spectral analysis}

Based on a short (4.7 ks) {\it Chandra} observation, it was suggested 
(Kaaret et al.~2003) that the X-ray spectrum of this ULX 
could be fitted equally well by a broken power law (break 
energy $\approx 0.65$ keV), or by an absorbed blackbody plus 
power-law model ($kT_{\rm bb} \approx 0.11$ keV; photon index 
$\Gamma \approx 2.8$). The broken power-law model would support 
the interpretation of this source as beamed emission via inverse-Compton 
scattering of optical/UV photons by a relativistic jet 
(Georganopoulos et al.~2001; Kaaret et al.~2003).

In our {\it XMM-Newton} study, we confirm that a simple absorbed 
power-law model is ruled out ($\chi^2_{\nu} \ga 1.4$ for all spectra).
We then consider a broken power-law model; we impose the physical 
constraint that the photon index below the break must be lower than 
the index above the break. Although at some epochs this model 
provides a good fit, with parameters similar to those of 
Kaaret et al.~(2003), at other epochs (e.g. for Rev.~305, 2001 Aug 08) 
it leads to large, systematic residuals (Tables 2, 3). 
Instead, a blackbody/disk-blackbody plus power-law model 
provides a consistently good fit over all the observations.
In this model, the power-law index $\Gamma \approx 2.6$--$2.9$ 
is similar to that found in Galactic BH candidates 
in the high/soft and very high states (Belloni 2001). 
The low-temperature thermal component 
is similar to the soft emission found in bright ULXs, 
such as those in NGC 1313 (Miller et al.~2003), Ho II 
(Dewangan et al.~2004) and NGC\,4559 (Cropper et al.~2004). We cannot 
discriminate from our observations between a pure blackbody 
and a multicolor disk-blackbody spectrum.
Comptonisation models such as {\tt {bmc}} (Shrader \& Titarchuk 1999) 
also provide a good fit (Tables 2, 3). 
%More complex spectral models, including the effect 
%of ionised or non-solar-abundance absorbers, will be discussed 
%in a follow-up paper. 
The thermal component is significantly detected at all epochs, 
with similar temperatures ($kT_{\rm bb} \approx 0.12$--$0.14$ keV).
We find long-term flux and spectral variations over the 18-month span 
of our observations (Fig.~1 and Tables 2, 3), but no real state 
transitions. At an assumed distance of 4.8 Mpc (Karachentsev et al.~2002), 
the emitted luminosity in the $0.3$--$12$ keV band is 
$\approx 10^{40}$ erg s$^{-1}$.

In the spectral models described above, we have assumed 
a solar metal abundance for the intrinsic absorber 
({\tt wabs} model in {\footnotesize{XSPEC}}). 
In fact, the metal abundance of the starburst region 
in NGC\,5408 is known to be lower, 
$Z \approx 1.2 \times 10^{-3} \approx 1/14\,Z_{\odot}$ (Stewart et al.~1982).
It is therefore likely that the intrinsic 
absorber in front of the ULX has also low metal abundance.
We have repeated the fits using a {\tt tbvarabs} model 
(Wilms et al.~2000) for the intrinsic absorber.
Firstly, we left the metal abundance as a free parameter: 
we found that its value is not well constrained, 
and we did not obtain significant $\chi^2$ improvements 
for any of the models. This is because the intrinsic 
column density is low, hence the fitted models are not 
much affected by changes in its metal abundance.
We then fixed the metal abundance at $Z = 0.07 Z_{\odot}$, 
and repeated the fits for the broken-power-law ({\tt{bknpo}}) 
and blackbody-plus-power-law ({\tt{bb+po}})
models (Tables 4, 5). Even with a lower metal abundance, 
including a low-temperature thermal component 
significantly improves the fits, compared with  
a simple power law. We also find once again that {\tt{bb+po}}
models give a better $\chi^2_\nu$ than {\tt{bknpo}} 
models, for 4 out of the 5 spectral observations, 
although we cannot rule out the broken-power-law model 
altogether.

We tried to find a quantitative way to estimate whether 
the {\tt{bb+po}} model is ``significantly'' better 
than the {\tt{bknpo}} model, as suggested by the differences in $\chi^2_\nu$ 
and null hypothesis probabilities, or whether the differences 
between the two can be attributed to chance.
Two models can be directly compared with the F-test (Hamilton 1965) 
in the special case when the parameters of one model are a subset 
of the parameters of the other model. 
In a more general case, such as the one we are dealing with, 
a quantitative test for comparing the predictions of two models 
with the observed values in each data bin was first discussed 
in Williams \& Kloot (1953), and refined in Himmelblau (1970) 
and Prince (1982). We summarize the results of our comparison 
in Table 6: in 4 of the 5 observations, there is a $\ga 90\%$ probability 
that the {\tt{bb+po}} model is a significantly better fit 
than the {\tt{bknpo}} model. Only for the last observation 
(2003 Jan) are the two models statistically indistinguishable.

A {\tt{bb+po}} model is also a significantly better fit than 
the {\tt{bknpo}} model for the average of the 3 observations 
from 2001 (Tables 2-5). This is not in itself a proof against 
the {\tt{bknpo}} model: it may simply result from 
the individual spectra being broken power-laws 
but with different parameters.  

In conclusion, we argue that a {\tt{bb+po}} model, or any other 
similar model with a power-law and a soft thermal component 
(eg, a Comptonisation model such as {\tt {bmc}}),  
are significantly better fits to the {\it XMM-Newton} data than 
either a simple or a broken power law.
This does not rule out a relativistic jet interpretation 
for this ULX: it was shown by Georganopoulos et al.~(2001) 
that a broken power-law spectrum is only an approximation, 
and that the true spectrum would in fact be curved around 
the break energy (in our case, $\approx 0.5$--$0.6$ keV). 
Such a spectral model would be practically indistinguishable 
from a power-law plus thermal component: observations 
over a larger energy range are needed to test this prediction.

%

%__________________________________________________ One column table
\begin{centering}
   \begin{table}
      \caption[]{Best-fit parameters for the combined EPIC spectra 
of the ULX during the three observations in 2001 (Rev~301: Jul 31; 
Rev.~305: Aug 08; Rev.~313: Aug 24). 
The quoted errors are the 90\% confidence limit for one parameter 
($\Delta \chi^2 = 2.7$). We assumed a Galactic column 
density $n_{\rm H,Gal} = 5.7 \times 10^{20}$ cm$^{-2}$ 
(Dickey \& Lockman 1990). Fluxes and luminosities are the unabsorbed 
values.}
         \label{table1}
\begin{centering}
         \begin{tabular}{l@{\hspace{0.2cm}}r@{\hspace{0.2cm}}r@{\hspace{0.2cm}}r}
            \hline
            \hline
            \noalign{\smallskip}
            Parameter      & Value Jul 31 &  Value Aug 08 & Value Aug 24 \\[2pt]
            \noalign{\smallskip}
	    \hline
            \hline
            \noalign{\smallskip}
            \noalign{\smallskip}
	\multicolumn{4}{c}{model: wabs$_{\rm Gal}~\times$ 
		wabs $\times$ bknpo}\\
            \noalign{\smallskip}
            \hline
            \noalign{\smallskip}
            \noalign{\smallskip}
%            	$n_{\rm H}~(10^{21}~{\rm cm}^{-2})$ & $1.9^{+0.2}_{-0.2}$ 
            	$n_{\rm H}~^{\mathrm{(a)}}$ & $0.2^{+0.3}_{-0.2}$
			& $0.2^{+0.3}_{-0.2}$ & $<0.3$     \\[2pt]
		$\Gamma_1$ & $0.98^{+0.81}_{-1.24}$ & $1.72^{+0.47}_{-0.58}$ 
				& $0.65^{+1.26}_{-1.44}$\\[2pt]
		$E_{\rm br}~({\rm keV})$ & $0.55^{+0.04}_{-0.04}$ 
			& $0.63^{+0.05}_{-0.10}$ & $0.53^{+0.13}_{-0.05}$\\[2pt]
		$\Gamma_2$  & $3.38^{+0.14}_{-0.13}$ & $3.34^{+0.13}_{-0.12}$ 
			& $3.15^{+0.15}_{-0.07}$\\[2pt]
		$K_{\rm bp}~(10^{-3})$ & $3.5^{+5.2}_{-1.4}$ & $1.7^{+2.7}_{-3.5}$
			& $3.3^{+5.5}_{-2.8}$\\
            \noalign{\smallskip}
            \hline
            \noalign{\smallskip}
		$\chi^2_\nu$ & $1.16\,(164.2/141)$ & $1.45\,(229.0/158)$ 
			& $0.99\,(132.2/133)$\\[2pt]
%            	$f_{0.3{\rm -}12}~(10^{-12}~{\rm erg~cm}^{-2}~{\rm s}^{-1})$ 
            	$f_{0.3{\rm -}12}~^{\mathrm{(b)}}$
			& $3.6^{+0.1}_{-1.2}$ & 
			$3.5^{+0.2}_{-1.0}$ & $3.2^{+0.4}_{-0.4}$\\[2pt]
%            	$L_{0.3{\rm -}12}~(10^{40}~{\rm erg~s}^{-1})$ & $4.5$ & 
            	$L_{0.3{\rm -}12}~^{\mathrm{(c)}}$ & $1.0$ &
			$1.0$ & $0.9$     \\
            \noalign{\smallskip}
	    \hline
            \noalign{\smallskip}
            \noalign{\smallskip}
	\multicolumn{4}{c}{model: wabs$_{\rm Gal}~\times$ 
		wabs $\times$ (bb $+$ po)}\\
            \noalign{\smallskip}
            \hline
            \noalign{\smallskip}
            \noalign{\smallskip}
%            	$n_{\rm H}~(10^{21}~{\rm cm}^{-2})$ & $0.4^{+0.1}_{-0.2}$ & 
            	$n_{\rm H}~^{\mathrm{(a)}}$ & $0.4^{+0.3}_{-0.2}$ &
			$0.3^{+0.2}_{-0.1}$ & $0.8^{+0.5}_{-0.3}$     \\[2pt]
		$kT_{\rm bb}~({\rm keV})$ & $0.14^{+0.01}_{-0.02}$ & 
			$0.15^{+0.01}_{-0.01}$ & $0.13^{+0.01}_{-0.02}$\\[2pt]
		$K_{\rm bb}~(10^{-5})$ & $2.5^{+0.9}_{-0.5}$ & $2.3^{+0.5}_{-0.3}$ 
			& $3.2^{+2.3}_{-1.0}$\\[2pt]
		$\Gamma$  & $2.91^{+0.20}_{-0.19}$ & $2.65^{+0.18}_{-0.18}$ & 
			$2.88^{+0.16}_{-0.18}$\\[2pt]
		$K_{\rm po}~(10^{-4})$ & $5.7^{+1.2}_{-1.1}$ & 
			$4.6^{+0.9}_{-0.8}$ & $6.7^{+1.3}_{-1.1}$\\
            \noalign{\smallskip}
            \hline
            \noalign{\smallskip}
		$\chi^2_\nu$ & $1.00\,(141.4/141)$ & $1.09\,(172.0/158)$ 
			& $0.86\,(113.9/133)$\\[2pt]
%            	$f_{0.3{\rm -}12}~(10^{-12}~{\rm erg~cm}^{-2}~{\rm s}^{-1})$ 
            	$f_{0.3{\rm -}12}~^{\mathrm{(b)}}$
			& $4.5^{+0.3}_{-0.9}$ & $3.8^{+0.2}_{-0.6}$ 
			& $5.5^{+0.3}_{-2.0}$\\[2pt]
%            	$L_{0.3{\rm -}12}~(10^{40}~{\rm erg~s}^{-1})$ & $1.2$ & $1.1$ 
            	$L_{0.3{\rm -}12}~^{\mathrm{(c)}}$ & $1.3$ & $1.1$
			& $1.5$  \\
            \noalign{\smallskip}
	    \hline
            \noalign{\smallskip}
            \noalign{\smallskip}
	\multicolumn{4}{c}{model: wabs$_{\rm Gal}~\times$ 
		wabs $\times$ (diskbb $+$ po)}\\
            \noalign{\smallskip}
            \hline
            \noalign{\smallskip}
            \noalign{\smallskip}
%		$n_{\rm H}~(10^{21}~{\rm cm}^{-2})$ & $0.7^{+0.2}_{-0.2}$ 
		$n_{\rm H}~^{\mathrm{(a)}}$ & $0.6^{+0.2}_{-0.4}$
			& $0.5^{+0.2}_{-0.2}$ & $1.1^{+0.5}_{-0.4}$\\[2pt]
		$kT_{\rm in}~({\rm keV})$ & $0.17^{+0.02}_{-0.02}$ 
			& $0.19^{+0.01}_{-0.02}$ & $0.15^{+0.03}_{-0.02}$\\[2pt]
		$K_{\rm dbb}$ & $277^{+390}_{-143}$ & $174^{+178}_{-48}$ 
			& $555^{+1580}_{-456}$\\[2pt]
		$\Gamma$  & $2.85^{+0.19}_{-0.20}$ & $2.55^{+0.24}_{-0.17}$ & 
			$2.86^{+0.19}_{-0.18}$\\[2pt]
		$K_{\rm po}~(10^{-4})$ & $5.3^{+1.3}_{-1.1}$ & 
			$4.1^{+1.2}_{-0.7}$ & $6.6^{+0.7}_{-0.6}$\\
            \noalign{\smallskip}
            \hline
            \noalign{\smallskip}
		$\chi^2_\nu$ & $0.99\,(139.6/141)$ & $1.05\,(166.2/158)$ 
			& $0.86\,(114.4/133)$\\[2pt]
%            	$f_{0.3{\rm -}12}~(10^{-12}~{\rm erg~cm}^{-2}~{\rm s}^{-1})$ 
            	$f_{0.3{\rm -}12}~^{\mathrm{(b)}}$
			& $5.2^{+0.1}_{-1.6}$ & $4.5^{+0.1}_{-0.9}$ 
			& $6.7^{+0.2}_{-2.8}$\\[2pt]
%            	$L_{0.3{\rm -}12}~(10^{40}~{\rm erg~s}^{-1})$ & $1.4$ & $1.4$ &$1.4$    \\
            	$L_{0.3{\rm -}12}~^{\mathrm{(c)}}$ & $1.4$ & $1.2$ &$1.8$    \\
            \noalign{\smallskip}
	    \hline
            \noalign{\smallskip}
            \noalign{\smallskip}
	\multicolumn{4}{c}{model: wabs$_{\rm Gal}~\times$ 
		wabs $\times$ bmc}\\
            \noalign{\smallskip}
            \hline
            \noalign{\smallskip}
            \noalign{\smallskip}
%		$n_{\rm H}~(10^{21}$~cm$^{-2})$ &$1.0^{+0.8}_{-1.0}$
		$n_{\rm H}~^{\mathrm{(a)}}$ &$<0.4$ 
			& $<0.2$ & $0.5^{+0.5}_{-0.5}$\\[2pt] 
		$kT_{\rm bb}$~(keV)   & $0.13^{+0.01}_{-0.01}$ 
			& $0.14^{+0.01}_{-0.01}$ & $0.12^{+0.02}_{-0.02}$\\[2pt]
		$\Gamma$         & $2.94^{+0.19}_{-0.19}$ & $2.68^{+0.12}_{-0.12}$ 
			& $2.90^{+0.22}_{-0.22}$\\[2pt]		
		log(A)         &$-0.18^{+0.15}_{-0.12}$ & $-0.30^{+0.12}_{-0.12}$ 
			& $-0.24^{+0.19}_{-0.16}$\\[2pt]
		$K_{\rm bmc}~(10^{-5})$ & $3.9^{+1.3}_{-0.2}$
			& $3.4^{+0.6}_{-0.1}$ & $5.3^{+2.4}_{-1.5}$\\
            \noalign{\smallskip}
            \hline
            \noalign{\smallskip}
		$\chi^2_\nu$ & $1.02\,(144.2/141)$ & $1.13\,(178.4/158)$ 
			& $0.86\,(114.8/133)$\\[2pt]
%            	$f_{0.3{\rm -}12}~(10^{-12}~{\rm erg~cm}^{-2}~{\rm s}^{-1})$ 
            	$f_{0.3{\rm -}12}~^{\mathrm{(b)}}$
			& $3.4^{+0.1}_{-0.8}$ & $3.1^{+0.4}_{-0.8}$ 
			& $4.4^{+1.1}_{-1.7}$\\[2pt]
%            	$L_{0.3{\rm -}12}~(10^{40}~{\rm erg~s}^{-1})$ & $0.9$ & $0.9$ & $0.9$   \\
            	$L_{0.3{\rm -}12}~^{\mathrm{(c)}}$ & $0.9$ & $0.9$ & $1.2$   \\
            \noalign{\smallskip}
	    \hline
         \end{tabular}
\begin{list}{}{}
%\item[$^{\mathrm{a}}$] This is footnote a
\item[$^{\mathrm{a}}$] in units of $10^{21}~{\rm cm}^{-2}$
\item[$^{\mathrm{b}}$] in units of $10^{-12}~{\rm erg~cm}^{-2}~{\rm s}^{-1}$
\item[$^{\mathrm{c}}$] in units of $10^{40}~{\rm erg~s}^{-1}$
\end{list}
\end{centering}
   \end{table}
\end{centering}

%__________________________________________________ One column table
\begin{centering}
   \begin{table}
      \caption[]{Best-fit parameters for the combined EPIC spectra 
of the ULX in 2001 (Jul 24--Aug 25), 2002 (Jul 29) and 2003 (Jan 28). 
The quoted errors are the 90\% confidence limit for one parameter 
($\Delta \chi^2 = 2.7$). We assumed a Galactic column 
density $n_{\rm H,Gal} = 5.7 \times 10^{20}$ cm$^{-2}$ 
(Dickey \& Lockman 1990). Fluxes and luminosities are the unabsorbed 
values.}
         \label{table1}
\begin{centering}
         \begin{tabular}{l@{\hspace{0.2cm}}r@{\hspace{0.2cm}}r@{\hspace{0.2cm}}r}
            \hline
            \hline
            \noalign{\smallskip}
            Parameter      & Value in 2001 &  Value in 2002 & Value in 2003 \\[2pt]
            \noalign{\smallskip}
	    \hline
            \hline
            \noalign{\smallskip}
            \noalign{\smallskip}
	\multicolumn{4}{c}{model: wabs$_{\rm Gal}~\times$ 
		wabs $\times$ bknpo}\\
            \noalign{\smallskip}
            \hline
            \noalign{\smallskip}
            \noalign{\smallskip}
%            	$n_{\rm H}~(10^{21}~{\rm cm}^{-2})$ & $1.9^{+0.2}_{-0.2}$ 
            	$n_{\rm H}~^{\mathrm{(a)}}$ & $0.1^{+0.1}_{-0.1}$
			& $<0.2$ & $<0.2$     \\[2pt]
		$\Gamma_1$ & $0.76^{+0.16}_{-0.48}$ & $0.58^{+0.52}_{-1.35}$ 
				& $-1.19^{+0.68}_{-1.75}$\\[2pt]
		$E_{\rm br}~({\rm keV})$ & $0.56^{+0.05}_{-0.03}$ 
			& $0.60^{+0.05}_{-0.06}$ & $0.55^{+0.05}_{-0.03}$\\[2pt]
		$\Gamma_2$  & $3.21^{+0.03}_{-0.02}$ & $2.89^{+0.11}_{-0.07}$ 
			& $2.87^{+0.06}_{-0.06}$\\[2pt]
		$K_{\rm bp}~(10^{-3})$ & $3.4^{+0.2}_{-0.2}$ & $2.0^{+3.3}_{-0.6}$
			& $6.2^{+0.6}_{-0.2}$\\
            \noalign{\smallskip}
            \hline
            \noalign{\smallskip}
		$\chi^2_\nu$ & $1.79\,(245.4/137)$ & $1.09\,(99.2/91)$ 
			& $0.97\,(83.2/86)$\\[2pt]
%            	$f_{0.3{\rm -}12}~(10^{-12}~{\rm erg~cm}^{-2}~{\rm s}^{-1})$ 
            	$f_{0.3{\rm -}12}~^{\mathrm{(b)}}$
			& $3.2^{+0.2}_{-0.2}$ & 
			$2.3^{+0.1}_{-0.1}$ & $2.0^{+0.1}_{-0.1}$\\[2pt]
%            	$L_{0.3{\rm -}12}~(10^{40}~{\rm erg~s}^{-1})$ & $4.5$ & 
            	$L_{0.3{\rm -}12}~^{\mathrm{(c)}}$ & $0.9$ &
			$0.6$ & $0.6$     \\
            \noalign{\smallskip}
	    \hline
            \noalign{\smallskip}
            \noalign{\smallskip}
	\multicolumn{4}{c}{model: wabs$_{\rm Gal}~\times$ 
		wabs $\times$ (bb $+$ po)}\\
            \noalign{\smallskip}
            \hline
            \noalign{\smallskip}
            \noalign{\smallskip}
%            	$n_{\rm H}~(10^{21}~{\rm cm}^{-2})$ & $0.4^{+0.1}_{-0.2}$ & 
            	$n_{\rm H}~^{\mathrm{(a)}}$ & $0.4^{+0.1}_{-0.2}$ &
			$0.9^{+0.1}_{-0.1}$ & $1.3^{+0.6}_{-0.5}$     \\[2pt]
		$kT_{\rm bb}~({\rm keV})$ & $0.14^{+0.01}_{-0.01}$ & 
			$0.13^{+0.01}_{-0.02}$ & $0.12^{+0.02}_{-0.02}$\\[2pt]
		$K_{\rm bb}~(10^{-5})$ & $2.5^{+0.5}_{-0.3}$ & $2.5^{+0.4}_{-0.5}$ 
			& $3.1^{+3.5}_{-1.3}$\\[2pt]
		$\Gamma$  & $2.79^{+0.11}_{-0.10}$ & $2.58^{+0.06}_{-0.06}$ & 
			$2.68^{+0.18}_{-0.17}$\\[2pt]
		$K_{\rm po}~(10^{-4})$ & $5.7^{+0.6}_{-0.6}$ & 
			$5.0^{+0.2}_{-0.3}$ & $4.9^{+1.0}_{-0.9}$\\
            \noalign{\smallskip}
            \hline
            \noalign{\smallskip}
		$\chi^2_\nu$ & $1.10\,(150.9/137)$ & $0.98\,(88.9/91)$ 
			& $0.99\,(85.5/86)$\\[2pt]
%            	$f_{0.3{\rm -}12}~(10^{-12}~{\rm erg~cm}^{-2}~{\rm s}^{-1})$ 
            	$f_{0.3{\rm -}12}~^{\mathrm{(b)}}$
			& $4.5^{+0.1}_{-0.5}$ & $4.0^{+0.1}_{-0.2}$ 
			& $4.3^{+0.2}_{-1.9}$\\[2pt]
%            	$L_{0.3{\rm -}12}~(10^{40}~{\rm erg~s}^{-1})$ & $1.2$ & $1.1$ 
            	$L_{0.3{\rm -}12}~^{\mathrm{(c)}}$ & $1.2$ & $1.1$
			& $1.2$  \\
            \noalign{\smallskip}
	    \hline
            \noalign{\smallskip}
            \noalign{\smallskip}
	\multicolumn{4}{c}{model: wabs$_{\rm Gal}~\times$ 
		wabs $\times$ (diskbb $+$ po)}\\
            \noalign{\smallskip}
            \hline
            \noalign{\smallskip}
            \noalign{\smallskip}
%		$n_{\rm H}~(10^{21}~{\rm cm}^{-2})$ & $0.7^{+0.2}_{-0.2}$ 
		$n_{\rm H}~^{\mathrm{(a)}}$ & $0.7^{+0.2}_{-0.2}$
			& $1.3^{+0.5}_{-0.4}$ & $1.5^{+0.7}_{-0.5}$\\[2pt]
		$kT_{\rm in}~({\rm keV})$ & $0.17^{+0.01}_{-0.01}$ 
			& $0.15^{+0.01}_{-0.01}$ & $0.15^{+0.02}_{-0.02}$\\[2pt]
		$K_{\rm dbb}$ & $273^{+165}_{-97}$ & $471^{+952}_{-289}$ 
			& $670^{+2300}_{-480}$\\[2pt]
		$\Gamma$  & $2.73^{+0.11}_{-0.11}$ & $2.57^{+0.19}_{-0.17}$ & 
			$2.66^{+0.18}_{-0.17}$\\[2pt]
		$K_{\rm po}~(10^{-4})$ & $5.3^{+0.7}_{-0.6}$ & 
			$5.0^{+0.9}_{-0.9}$ & $4.8^{+1.1}_{-0.8}$\\
            \noalign{\smallskip}
            \hline
            \noalign{\smallskip}
		$\chi^2_\nu$ & $1.05\,(144.5/137)$ & $1.02\,(93.1/91)$ 
			& $1.02\,(87.4/86)$\\[2pt]
%            	$f_{0.3{\rm -}12}~(10^{-12}~{\rm erg~cm}^{-2}~{\rm s}^{-1})$ 
            	$f_{0.3{\rm -}12}~^{\mathrm{(b)}}$
			& $5.2^{+0.2}_{-0.8}$ & $5.0^{+2.0}_{-1.0}$ 
			& $5.2^{+4.0}_{-1.5}$\\[2pt]
%            	$L_{0.3{\rm -}12}~(10^{40}~{\rm erg~s}^{-1})$ & $1.4$ & $1.4$ &$1.4$    \\
            	$L_{0.3{\rm -}12}~^{\mathrm{(c)}}$ & $1.4$ & $1.4$ &$1.4$    \\
            \noalign{\smallskip}
	    \hline
            \noalign{\smallskip}
            \noalign{\smallskip}
	\multicolumn{4}{c}{model: wabs$_{\rm Gal}~\times$ 
		wabs $\times$ bmc}\\
            \noalign{\smallskip}
            \hline
            \noalign{\smallskip}
            \noalign{\smallskip}
%		$n_{\rm H}~(10^{21}$~cm$^{-2})$ &$1.0^{+0.8}_{-1.0}$
		$n_{\rm H}~^{\mathrm{(a)}}$ &$1.0^{+0.8}_{-1.0}$ 
			& $0.6^{+0.5}_{-0.2}$ & $0.9^{+0.4}_{-0.4}$\\[2pt] 
		$kT_{\rm bb}$~(keV)   & $0.13^{+0.02}_{-0.01}$ 
			& $0.13^{+0.02}_{-0.02}$ & $0.12^{+0.02}_{-0.02}$\\[2pt]
		$\Gamma$         & $2.78^{+0.12}_{-0.10}$ & $2.60^{+0.07}_{-0.07}$ 
			& $2.67^{+0.20}_{-0.17}$\\[2pt]		
		log(A)         &$-0.20^{+0.10}_{-0.8}$ & $-0.24^{+0.04}_{-0.04}$ 
			& $-0.30^{+0.20}_{-0.20}$\\[2pt]
		$K_{\rm bmc}~(10^{-5})$ & $3.8^{+0.7}_{-0.2}$
			& $3.6^{+0.3}_{-0.2}$ & $4.0^{+3.5}_{-2.0}$\\
            \noalign{\smallskip}
            \hline
            \noalign{\smallskip}
		$\chi^2_\nu$ & $1.17\,(160.5/137)$ & $0.95\,(86.4/91)$ 
			& $0.97\,(83.5/86)$\\[2pt]
%            	$f_{0.3{\rm -}12}~(10^{-12}~{\rm erg~cm}^{-2}~{\rm s}^{-1})$ 
            	$f_{0.3{\rm -}12}~^{\mathrm{(b)}}$
			& $3.5^{+0.3}_{-0.9}$ & $3.3^{+0.4}_{-1.4}$ 
			& $3.4^{+0.4}_{-1.6}$\\[2pt]
%            	$L_{0.3{\rm -}12}~(10^{40}~{\rm erg~s}^{-1})$ & $0.9$ & $0.9$ & $0.9$   \\
            	$L_{0.3{\rm -}12}~^{\mathrm{(c)}}$ & $1.0$ & $0.9$ & $0.9$   \\
            \noalign{\smallskip}
	    \hline
         \end{tabular}
\begin{list}{}{}
%\item[$^{\mathrm{a}}$] This is footnote a
\item[$^{\mathrm{a}}$] in units of $10^{21}~{\rm cm}^{-2}$
\item[$^{\mathrm{b}}$] in units of $10^{-12}~{\rm erg~cm}^{-2}~{\rm s}^{-1}$
\item[$^{\mathrm{c}}$] in units of $10^{40}~{\rm erg~s}^{-1}$
\end{list}
\end{centering}
   \end{table}
\end{centering}

%__________________________________________________ One column table
\begin{centering}
   \begin{table}
      \caption[]{As in Table 2, but with sub-solar abundances 
($Z = 0.07 Z_{\odot}$) for the intrinsic absorber.}
         \label{table1}
\begin{centering}
         \begin{tabular}{l@{\hspace{0.2cm}}r@{\hspace{0.2cm}}r@{\hspace{0.2cm}}r}
            \hline
            \hline
            \noalign{\smallskip}
            Parameter      & Value Jul 31 &  Value Aug 08 & Value Aug 24 \\[2pt]
            \noalign{\smallskip}
	    \hline
            \hline
            \noalign{\smallskip}
            \noalign{\smallskip}
	\multicolumn{4}{c}{model: wabs$_{\rm Gal}~\times$ 
		tbvarabs $\times$ bknpo}\\
            \noalign{\smallskip}
            \hline
            \noalign{\smallskip}
            \noalign{\smallskip}
%            	$n_{\rm H}~(10^{21}~{\rm cm}^{-2})$ & $1.9^{+0.2}_{-0.2}$ 
            	$n_{\rm H}~^{\mathrm{(a)}}$ & $0.6^{+0.4}_{-0.4}$
			& $0.9^{+0.4}_{-0.7}$ & $0.2^{+0.7}_{-0.2}$     \\[2pt]
		$\Gamma_1$ & $1.61^{+1.30}_{-1.52}$ & $2.12^{+1.68}_{-\ast}$ 
				& $1.02^{+1.60}_{-1.54}$\\[2pt]
		$E_{\rm br}~({\rm keV})$ & $0.53^{+0.06}_{-0.06}$ 
			& $0.51^{+0.35}_{-0.20}$ & $0.57^{+0.13}_{-0.13}$\\[2pt]
		$\Gamma_2$  & $3.40^{+0.14}_{-0.13}$ & $3.38^{+0.10}_{-0.11}$ 
			& $3.18^{+0.15}_{-0.07}$\\[2pt]
		$K_{\rm bp}~(10^{-3})$ & $2.7^{+5.8}_{-1.5}$ & $2.0^{+\ast}_{-\ast}$
			& $2.8^{+\ast}_{-1.0}$\\
            \noalign{\smallskip}
            \hline
            \noalign{\smallskip}
		$\chi^2_\nu$ & $1.15\,(162.3/141)$ & $1.43\,(225.3/158)$ 
			& $0.99\,(132.0/133)$\\[2pt]
%            	$f_{0.3{\rm -}12}~(10^{-12}~{\rm erg~cm}^{-2}~{\rm s}^{-1})$ 
            	$f_{0.3{\rm -}12}~^{\mathrm{(b)}}$
			& $4.0^{+0.8}_{-0.7}$ & 
			$4.4^{+1.0}_{-1.1}$ & $3.4^{+1.0}_{-0.2}$\\[2pt]
%            	$L_{0.3{\rm -}12}~(10^{40}~{\rm erg~s}^{-1})$ & $4.5$ & 
            	$L_{0.3{\rm -}12}~^{\mathrm{(c)}}$ & $1.1$ &
			$1.2$ & $0.9$     \\
            \noalign{\smallskip}
	    \hline
            \noalign{\smallskip}
            \noalign{\smallskip}
	\multicolumn{4}{c}{model: wabs$_{\rm Gal}~\times$ 
		tbvarabs $\times$ (bb $+$ po)}\\
            \noalign{\smallskip}
            \hline
            \noalign{\smallskip}
            \noalign{\smallskip}
%            	$n_{\rm H}~(10^{21}~{\rm cm}^{-2})$ & $0.4^{+0.1}_{-0.2}$ & 
            	$n_{\rm H}~^{\mathrm{(a)}}$ & $0.5^{+0.2}_{-0.2}$ &
			$0.3^{+0.2}_{-0.2}$ & $0.7^{+0.2}_{-0.3}$     \\[2pt]
		$kT_{\rm bb}~({\rm keV})$ & $0.14^{+0.01}_{-0.01}$ & 
			$0.15^{+0.01}_{-0.01}$ & $0.13^{+0.01}_{-0.02}$\\[2pt]
		$K_{\rm bb}~(10^{-5})$ & $1.9^{+0.4}_{-0.5}$ & $2.0^{+0.2}_{-0.3}$ 
			& $1.9^{+1.3}_{-0.9}$\\[2pt]
		$\Gamma$  & $2.93^{+0.17}_{-0.23}$ & $2.61^{+0.18}_{-0.18}$ & 
			$2.85^{+0.16}_{-0.18}$\\[2pt]
		$K_{\rm po}~(10^{-4})$ & $5.7^{+0.9}_{-1.2}$ & 
			$4.4^{+0.9}_{-0.8}$ & $6.4^{+1.0}_{-1.0}$\\
            \noalign{\smallskip}
            \hline
            \noalign{\smallskip}
		$\chi^2_\nu$ & $1.00\,(141.6/141)$ & $1.10\,(173.2/158)$ 
			& $0.87\,(115.2/133)$\\[2pt]
%            	$f_{0.3{\rm -}12}~(10^{-12}~{\rm erg~cm}^{-2}~{\rm s}^{-1})$ 
            	$f_{0.3{\rm -}12}~^{\mathrm{(b)}}$
			& $4.1^{+0.5}_{-0.4}$ & $3.5^{+0.3}_{-0.2}$ 
			& $4.4^{+0.5}_{-0.5}$\\[2pt]
%            	$L_{0.3{\rm -}12}~(10^{40}~{\rm erg~s}^{-1})$ & $1.2$ & $1.1$ 
            	$L_{0.3{\rm -}12}~^{\mathrm{(c)}}$ & $1.1$ & $1.0$
			& $1.2$  \\
            \noalign{\smallskip}
	    \hline
         \end{tabular}
\begin{list}{}{}
%\item[$^{\mathrm{a}}$] This is footnote a
\item[$^{\mathrm{a}}$] in units of $10^{21}~{\rm cm}^{-2}$
\item[$^{\mathrm{b}}$] in units of $10^{-12}~{\rm erg~cm}^{-2}~{\rm s}^{-1}$
\item[$^{\mathrm{c}}$] in units of $10^{40}~{\rm erg~s}^{-1}$
\end{list}
\end{centering}
   \end{table}
\end{centering}

%__________________________________________________ One column table
\begin{centering}
   \begin{table}
      \caption[]{As in Table 3, but with sub-solar abundances 
($Z = 0.07 Z_{\odot}$) for the intrinsic absorber.}
         \label{table1}
\begin{centering}
         \begin{tabular}{l@{\hspace{0.2cm}}r@{\hspace{0.2cm}}r@{\hspace{0.2cm}}r}
            \hline
            \hline
            \noalign{\smallskip}
            Parameter      & Value in 2001 &  Value in 2002 & Value in 2003 \\[2pt]
            \noalign{\smallskip}
	    \hline
            \hline
            \noalign{\smallskip}
            \noalign{\smallskip}
	\multicolumn{4}{c}{model: wabs$_{\rm Gal}~\times$ 
		tbvarabs $\times$ bknpo}\\
            \noalign{\smallskip}
            \hline
            \noalign{\smallskip}
            \noalign{\smallskip}
%            	$n_{\rm H}~(10^{21}~{\rm cm}^{-2})$ & $1.9^{+0.2}_{-0.2}$ 
            	$n_{\rm H}~^{\mathrm{(a)}}$ & $0.1^{+0.1}_{-0.1}$
			& $<0.2$ & $<0.2$     \\[2pt]
		$\Gamma_1$ & $0.76^{+0.16}_{-0.48}$ & $0.58^{+0.52}_{-1.35}$ 
				& $-1.19^{+0.68}_{-1.75}$\\[2pt]
		$E_{\rm br}~({\rm keV})$ & $0.55^{+0.05}_{-0.03}$ 
			& $0.60^{+0.05}_{-0.06}$ & $0.55^{+0.05}_{-0.03}$\\[2pt]
		$\Gamma_2$  & $3.20^{+0.03}_{-0.02}$ & $2.89^{+0.11}_{-0.07}$ 
			& $2.87^{+0.06}_{-0.06}$\\[2pt]
		$K_{\rm bp}~(10^{-3})$ & $3.4^{+0.2}_{-0.2}$ & $2.0^{+3.3}_{-0.6}$
			& $6.2^{+0.6}_{-0.2}$\\
            \noalign{\smallskip}
            \hline
            \noalign{\smallskip}
		$\chi^2_\nu$ & $1.78\,(244.0/137)$ & $1.09\,(99.2/91)$ 
			& $0.97\,(83.2/86)$\\[2pt]
%            	$f_{0.3{\rm -}12}~(10^{-12}~{\rm erg~cm}^{-2}~{\rm s}^{-1})$ 
            	$f_{0.3{\rm -}12}~^{\mathrm{(b)}}$
			& $3.2^{+0.2}_{-0.2}$ & 
			$2.3^{+0.1}_{-0.1}$ & $2.0^{+0.1}_{-0.1}$\\[2pt]
%            	$L_{0.3{\rm -}12}~(10^{40}~{\rm erg~s}^{-1})$ & $4.5$ & 
            	$L_{0.3{\rm -}12}~^{\mathrm{(c)}}$ & $0.9$ &
			$0.6$ & $0.6$     \\
            \noalign{\smallskip}
	    \hline
            \noalign{\smallskip}
            \noalign{\smallskip}
	\multicolumn{4}{c}{model: wabs$_{\rm Gal}~\times$ 
		tbvarabs $\times$ (bb $+$ po)}\\
            \noalign{\smallskip}
            \hline
            \noalign{\smallskip}
            \noalign{\smallskip}
%            	$n_{\rm H}~(10^{21}~{\rm cm}^{-2})$ & $0.4^{+0.1}_{-0.2}$ & 
            	$n_{\rm H}~^{\mathrm{(a)}}$ & $0.4^{+0.2}_{-0.2}$ &
			$1.0^{+0.1}_{-0.1}$ & $1.4^{+0.7}_{-0.4}$     \\[2pt]
		$kT_{\rm bb}~({\rm keV})$ & $0.15^{+0.01}_{-0.02}$ & 
			$0.13^{+0.01}_{-0.02}$ & $0.12^{+0.02}_{-0.02}$\\[2pt]
		$K_{\rm bb}~(10^{-5})$ & $1.8^{+0.2}_{-0.2}$ & $1.3^{+0.2}_{-0.2}$ 
			& $1.6^{+1.7}_{-0.6}$\\[2pt]
		$\Gamma$  & $2.75^{+0.11}_{-0.10}$ & $2.61^{+0.06}_{-0.05}$ & 
			$2.67^{+0.18}_{-0.17}$\\[2pt]
		$K_{\rm po}~(10^{-4})$ & $5.3^{+0.5}_{-0.5}$ & 
			$5.0^{+0.5}_{-0.2}$ & $4.8^{+1.0}_{-0.9}$\\
            \noalign{\smallskip}
            \hline
            \noalign{\smallskip}
		$\chi^2_\nu$ & $1.12\,(153.6/137)$ & $0.93\,(84.3/91)$ 
			& $0.91\,(78.3/86)$\\[2pt]
%            	$f_{0.3{\rm -}12}~(10^{-12}~{\rm erg~cm}^{-2}~{\rm s}^{-1})$ 
            	$f_{0.3{\rm -}12}~^{\mathrm{(b)}}$
			& $3.9^{+0.2}_{-0.3}$ & $3.3^{+0.2}_{-0.1}$ 
			& $3.3^{+0.9}_{-0.4}$\\[2pt]
%            	$L_{0.3{\rm -}12}~(10^{40}~{\rm erg~s}^{-1})$ & $1.2$ & $1.1$ 
            	$L_{0.3{\rm -}12}~^{\mathrm{(c)}}$ & $1.1$ & $0.9$
			& $0.9$  \\
            \noalign{\smallskip}
	    \hline
         \end{tabular}
\begin{list}{}{}
%\item[$^{\mathrm{a}}$] This is footnote a
\item[$^{\mathrm{a}}$] in units of $10^{21}~{\rm cm}^{-2}$
\item[$^{\mathrm{b}}$] in units of $10^{-12}~{\rm erg~cm}^{-2}~{\rm s}^{-1}$
\item[$^{\mathrm{c}}$] in units of $10^{40}~{\rm erg~s}^{-1}$
\end{list}
\end{centering}
   \end{table}
\end{centering}

%__________________________________________________ One column table
\begin{centering}
   \begin{table}
      \caption[]{Comparison between the absorbed broken power-law model 
fits (``model 1'') and the absorbed blackbody plus power law fits 
(``model 2''), over the 5 observations. $P^N_1$ and $P^N_2$ are 
the null hypothesis probabilities for the two models. $p_2$ is 
the probability that model 2 is significantly better than model 1, 
from the Williams \& Kloot (1953) test. (For the 2003 observation, 
neither model is significantly better than the other.)}
         \label{table1}
\begin{centering}
         \begin{tabular}{l@{\hspace{0.2cm}}c@{\hspace{0.2cm}}c@{\hspace{0.2cm}}c@{\hspace{0.2cm}}c@{\hspace{0.2cm}}c}
            \hline
            \hline
            \noalign{\smallskip}
            Fit stats      & 31/07/01 &  08/08/01 & 24/08/01 & 29/07/02 & 28/01/03\\[2pt]
            \noalign{\smallskip}
	    \hline
            \hline
            \noalign{\smallskip}
            \noalign{\smallskip}
	\multicolumn{6}{c}{$Z = Z_{\odot}$}\\
            \noalign{\smallskip}
            \hline
            \noalign{\smallskip}
            \noalign{\smallskip}
			$\chi^2_{\nu,\,1}$ & 1.16 & 1.45 & 0.99 & 1.09 & 0.97 \\[2pt]
			$\chi^2_{\nu,\,2}$ & 1.00 & 1.09 & 0.86 & 0.98 & 0.99 \\
            \noalign{\smallskip}
            \hline
            \noalign{\smallskip}
			$P^N_{1}$ &0.09 & $1.9\times10^{-4}$& 0.50 & 0.26 & 0.57\\[2pt]
			$P^N_{2}$ & 0.47 & 0.21& 0.88 & 0.54 & 0.50\\	
            \noalign{\smallskip}
            \hline
            \noalign{\smallskip}
			$p_{2}$ & $93.1\%$ & $>99.99\%$ & $96.8\%$ & $87.5\%$ & ---\\
            \noalign{\smallskip}
	    \hline
            \noalign{\smallskip}
            \noalign{\smallskip}
	\multicolumn{6}{c}{$Z = 0.07\,Z_{\odot}$}\\
            \noalign{\smallskip}
            \hline
            \noalign{\smallskip}
            \noalign{\smallskip}
			$\chi^2_{\nu,\,1}$ & 1.15 & 1.43 & 0.99 & 1.09 & 0.97 \\[2pt]
			$\chi^2_{\nu,\,2}$ & 1.00 & 1.10 & 0.87 & 0.93 & 0.91 \\
            \noalign{\smallskip}
            \hline
            \noalign{\smallskip}
			$P^N_{1}$ & 0.11 & $3.5\times10^{-4}$ & 0.51 & 0.26 & 0.57\\[2pt]
			$P^N_{2}$ & 0.47 & 0.19 & 0.86 & 0.68 & 0.71 \\	
            \noalign{\smallskip}
            \hline
            \noalign{\smallskip}
			$p_{2}$ & $93.9\%$ &  $99.9\%$ & $97.4\%$ & $99.4\%$& ---\\				
            \noalign{\smallskip}
	    \hline
         \end{tabular}
\end{centering}
   \end{table}
\end{centering}

%----------------------------------------------------------- S_vib
   \begin{figure}
%   \centering
%  \includegraphics[angle=-90, width=8.8cm]{epicall_bbpo.ps}
  \includegraphics[angle=-90, width=8.8cm]{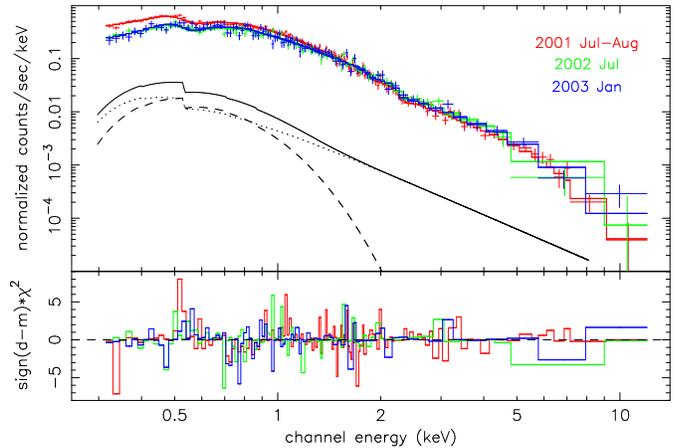}
      \caption{Coadded EPIC spectra of the ULX in 2001 Jul--Aug (blue, 
with its best-fit blackbody plus powerlaw model), 
2002 July (red), 2003 Jan (green). Also overplotted: the best-fit 
model for the 2001 spectrum, 
%(dashed line: blackbody; dotted line: power law), 
in photons cm$^{-1}$ s$^{-1}$ keV$^{-1}$ (arbitrarily shifted for clarity).
              }
         \label{Fig1}
   \end{figure}
%
%______________________________________________________________

%----------------------------------------------------------- S_vib
   \begin{figure}
%   \centering
%  \includegraphics[angle=-90, width=8.8cm]{epicall_bbpo.ps}
  \includegraphics[angle=-90, width=8.8cm]{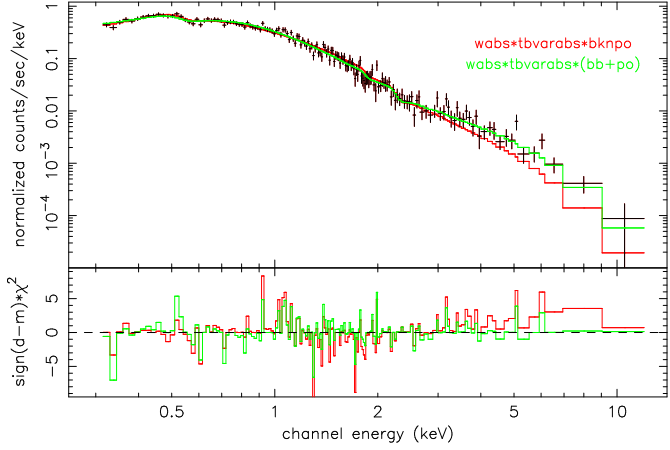}
      \caption{Coadded EPIC spectra of the ULX on 2001 Aug 08, 
with the best-fit {\tt {bb+po}} and {\tt {bknpo}} models 
(green and red curves, respectively). A metal abundance $Z = 0.07\,Z_{\odot}$ 
was assumed for the intrinsic absorber. See Table 4 for the fit parameters.       }
         \label{Fig1}
   \end{figure}
%
%______________________________________________________________

\subsection{Timing analysis}

Short-term variability, significant to $> 99$\% according to 
the Kolmogorov-Smirnoff and $\chi^2$ tests, is detected 
at all epochs, except during Rev.~313 (which has a lower 
signal-to-noise ratio, since the pn camera was not operating). 
The source exhibits a flaring behaviour, 
with changes in flux by a factor of $\sim 2$ over $\sim 100$ s (Fig.~3). 
The flares are most evident in the 2003 observation, with a 
Kolmogorov-Smirnoff probability of constancy $< 10^{-3}$.  
Here we focus on two important results from this epoch; 
a detailed timing study of all the other observations 
will be presented elsewhere.

Comparing the background-subtracted 
lightcurves in a soft ($0.2$--$1.5$ keV) and hard ($1.5$--$12$ keV) 
band, we found that the hard flux drops to a value consistent 
with zero during the dips (Fig.~4, top); in the soft band, 
a ``baseline'', persistent component is always present in addition 
to the flaring component (Fig.~4, middle). The rms fractional 
variability is $\approx 70\%$ in the hard band, and $\approx 30\%$ 
in the soft band. There is no time lag between the flares in the two bands, 
suggesting that we are seeing the same variable physical component 
in both. 

% I guess that the
%reader (and myself) would like to know whether there is any reasonable
%physical interpretation of the flaring spectral behaviour in the framework
%of the broken powerlaw model (considered here as an approximation of the
%true jet spectrum). Unless I missed it, you do not seem to consider in the
%conclusions that the clean spectral behaviour of the flares in the bb+pl
%scenario is an argument for this interpretation. I think that this ought
%to be mentioned somewhere.

The flaring behaviour in the two bands has 
a simple interpretation in the framework 
of the {\tt {bb+po}} spectral model (Sect.~3.1). 
In the hard band, the variability is necessarily due to the power-law component:
there is no contribution from the thermal component in this band.
A change in the power-law component would also affect 
the soft band, where the power-law and thermal fluxes 
are of the same order of magnitude.
Hence, we suggest that the lightcurves can be explained by 
a flaring behaviour of the power-law component, while the thermal  
component does not vary on these short timescales, 
and represents the baseline flux seen in the soft band.

To test our hypothesis, we analysed the lightcurve 
in the $0.2$--$12$ keV band (Fig.~3, bottom): 
we divided the exposure into a ``low'' interval, consisting 
of all the 50-s bins in which 
the count rate was $< 1$ ct s$^{-1}$, and a ``high'' interval, 
for rates $> 1$ ct s$^{-1}$. We then extracted and analysed 
the combined EPIC spectra from the two sub-intervals; as before, only 
the $0.3$--$12$ keV band was used for spectral 
fitting\footnote{The pn and MOS spectra were restricted to the 
same good-time-interval, for consistency with the lightcurve 
analysis. However, the exposure time is slightly longer 
in the MOS cameras (3.3 ks versus 3.0 ks for the pn, Table 1) 
because of the higher live-time fraction ($\approx 99$\% 
of the good-time-interval for the MOS CCDs, 
as opposed to $\approx 91$\% for the pn CCDs).}.
This choice of threshold is purely arbitrary: we needed 
to have enough counts in both ``states'' for a meaningful 
spectral analysis. We examined whether 
the change in flux between the peaks and the troughs 
could simply be modelled with a constant factor, 
or whether there was also a significant spectral 
change. We find (Fig.~6; Tables 7, 8) that the blackbody component 
is consistent with being constant between the two states:  
the flux variations are consistent with being 
entirely due to a change in the power-law normalisation, 
in agreement with our initial hypothesis. In conclusion, we suggest  
that the ULX exhibits soft dips over timescales of $\sim 10^2$ s 
in which the power-law component is suppressed.

We also fitted the two flux-dependent spectra 
with a {\tt{bknpo}} model (Table 9).
The model provides an equally good fit for this 
particular observations, as shown earlier (Tables 3, 5, 6).
However, its physical interpretation is less 
straightforward. It requires a very high power-law index below the break 
($\Gamma_1 \approx -1$ below $E_{\rm br} \approx 0.55$ keV), thus making 
the soft component practically indistinguishable from 
a thermal component over the {\it XMM-Newton}/EPIC energy band.
In this scenario, the flaring behaviour would correspond 
to a flattening of the high-energy power-law: from 
$\Gamma_2 \approx -3.2$ in the low state to 
$\Gamma_2 \approx -2.8$ in the high state
Finally, we show (Fig.~7) that the difference 
between low and high states cannot be due 
simply to a normalisation factor. The low state 
is indeed softer (consistent with our interpretation).

To investigate further the characteristic variability 
timescales, we examined the power spectral density 
for the 2003 lightcurve; we normalised the Fourier transform 
to be the squared rms fractional variability per unit frequency. 
Firstly, we considered in our analysis only the first uninterrupted 
3 ks of the exposure (Fig.~4); then we considered the power spectrum 
for the whole exposure, filling the two data gaps with a running mean. 
We obtained similar results.
The power spectrum for the full-band lightcurve (Fig.~5) has a slope 
$\alpha = -1.3 \pm 0.2$ over 
two decades in frequency ($3 \times 10^{-3} \la \nu \la 0.3$ Hz); 
it peaks at $\nu_{\rm b} \approx 2.5 \times 10^{-3}$ Hz, 
and is flat (consistent with $\alpha = 0$) at lower frequencies. 
The power spectrum for the hard-band lightcurve has similar 
features, with a break at $\nu_{\rm b} \approx 3 \times 10^{-3}$ Hz 
and a slope $\alpha = -0.9 \pm 0.4$ at higher frequencies.

%---------------------------------------------------------------
   \begin{figure}
%   \centering
  \includegraphics[angle=-90, width=8.8cm]{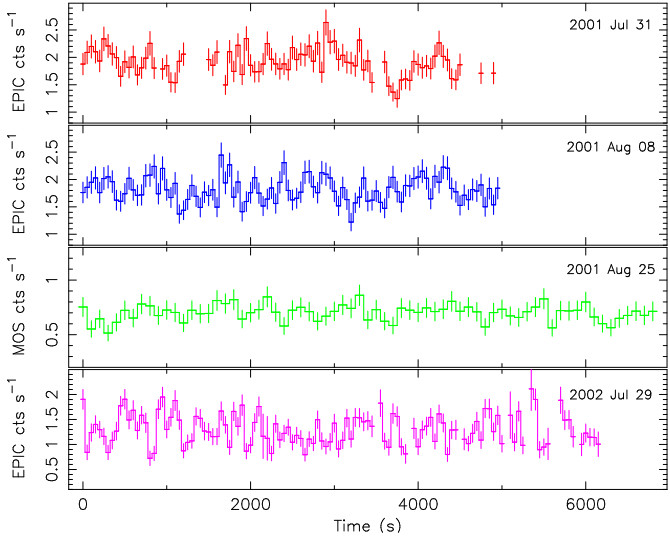}
      \caption{Coadded EPIC lightcurves of the ULX over 
the four observations of 2001--2002.}
%, grouped to 50-s bins 
%(100-s bins for the Rev.~313 data).}
         \label{Fig2}
   \end{figure}

%---------------------------------------------------------------
   \begin{figure}
%   \centering
  \includegraphics[angle=-90, width=8.8cm]{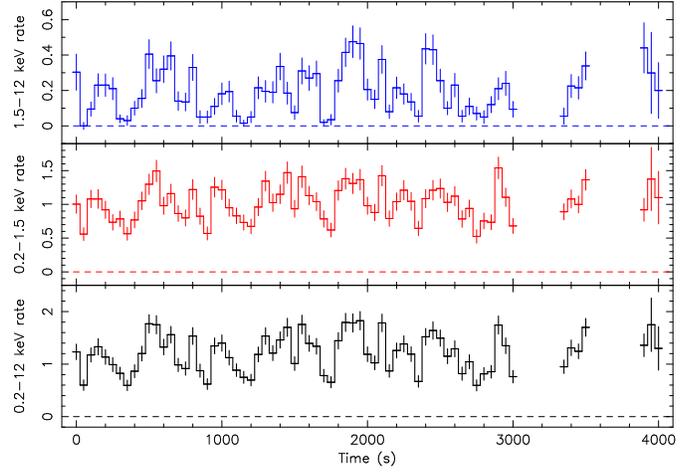}
      \caption{Coadded EPIC lightcurves of the ULX in 2003 Jan (Rev.~574), 
rebinned to 50 s. Top panel: 
$1.5$--$12$ keV band; medium panel: $0.2$--$1.5$ keV band; 
bottom panel: full $0.2$--$12$ keV band. 
              }
         \label{Fig3}
   \end{figure}

%----------------------------------------------------------- S_vib
   \begin{figure}
%   \centering
  \includegraphics[angle=-90, width=8.8cm]{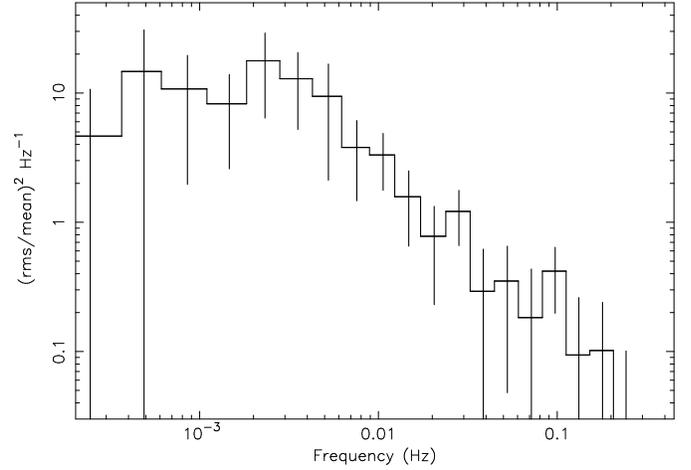}
      \caption{Power density spectrum of the source  
in 2003 Jan, normalised to give the squared rms fractional  
variability per unit frequency; the expected white noise 
level has been subtracted. Only the first, uninterrupted 
3 ks of the exposure have been used for this spectrum; 
however, very similar results are obtained when we used 
the whole exposure.
              }
         \label{Fig4}
   \end{figure}
%
%______________________________________________________________

\begin{centering}
   \begin{table}
      \caption[]{Best-fit parameters for a {\tt {bb+po}} simultaneous fit 
to the combined EPIC spectra in the ``low'' and ``high'' intervals 
of the 2003 Jan observation. As before, 
the quoted errors are the 90\% confidence limit and 
$n_{\rm H,Gal} = 5.7 \times 10^{20}$ cm$^{-2}$. We assumed a solar abundance 
for the intrinsic absorber ($Z = Z_{\odot}$).}
         \label{table1}
\begin{centering}
         \begin{tabular}{lrr}
            \hline
            \hline
            \noalign{\smallskip}
            Parameter    & ``Low'' value & ``High'' value   \\[2pt]
            \noalign{\smallskip}
	    \hline
            \hline
            \noalign{\smallskip}
            \noalign{\smallskip}
	\multicolumn{3}{c}{model: wabs$_{\rm Gal}~\times$ 
		wabs $\times$ (bb $+$ po)}\\
            \noalign{\smallskip}
            \hline
            \noalign{\smallskip}
            \noalign{\smallskip}
            	$n_{\rm H}~(10^{21}~{\rm cm}^{-2})$ & 
			$(1.4^{+0.5}_{-0.3})$ 
			& $(1.4^{+0.5}_{-0.3})~^{\mathrm{(a)}}$ \\[2pt]
		$kT_{\rm bb}~({\rm keV})$ &  $0.11^{+0.02}_{-0.01}$
			& $0.12^{+0.01}_{-0.02}$ \\[2pt]
		$K_{\rm bb}~(10^{-5})$ & $3.7^{+3.1}_{-0.9}$ 
			& $3.9^{+4.8}_{-1.0}$\\[2pt]
		$\Gamma$  &  $2.86^{+0.42}_{-0.37}$ & 
			$2.68^{+0.19}_{-0.18}$\\[2pt]
		$K_{\rm po}~(10^{-4})$ & 
			$3.1^{+1.0}_{-0.9}$ & $6.4^{+1.2}_{-1.2}$\\
            \noalign{\smallskip}
            \hline
            \noalign{\smallskip}
		$\chi^2_\nu$ & \multicolumn{2}{c}{~~~~~~~$0.76\,(137.8/182)$}  \\[2pt] 
		$f_{0.3{\rm -}12{\rm ,\,obs}}
			~(10^{-12}~{\rm erg~cm}^{-2}~{\rm s}^{-1})$
			&  $1.1^{+0.2}_{-0.4}$ & $2.0^{+0.2}_{-0.4}$\\[2pt] 
		$f_{0.3{\rm -}12}~(10^{-12}~{\rm erg~cm}^{-2}~{\rm s}^{-1})$
			&  $3.6^{+0.4}_{-1.9}$ & $5.5^{+0.3}_{-2.3}$\\[2pt]
		$f^{\mathrm{bb}}_{0.3{\rm -}12}~
			(10^{-12}~{\rm erg~cm}^{-2}~{\rm s}^{-1})$
			& $2.1^{+0.7}_{-1.6}$ & $2.3^{+1.1}_{-2.0}$  \\[2pt]
		$f^{\mathrm{po}}_{0.3{\rm -}12}~
			(10^{-12}~{\rm erg~cm}^{-2}~{\rm s}^{-1})$
			&  $1.5^{+1.2}_{-0.4}$ & $3.1^{+1.3}_{-0.6}$\\[2pt]
            	$L_{0.3{\rm -}12}~(10^{40}~{\rm erg~s}^{-1})$ & $1.0$ & $1.8$\\
            \noalign{\smallskip}
	    \hline
         \end{tabular}
\begin{list}{}{}
\item[$^{\mathrm{a}}$] assumed to be equal for the two states
%\item[$^{\mathrm{b}}$] in units of $10^{-12}~{\rm erg~cm}^{-2}~{\rm s}^{-1}$
%\item[$^{\mathrm{c}}$] in units of $10^{40}~{\rm erg~s}^{-1}$
\end{list}
\end{centering}
   \end{table}
\end{centering}

\begin{centering}
   \begin{table}
      \caption[]{As in Table 7, but with $Z = 0.07\,Z_{\odot}$.}
         \label{table1}
\begin{centering}
         \begin{tabular}{lrr}
            \hline
            \hline
            \noalign{\smallskip}
            Parameter    & ``Low'' value & ``High'' value   \\[2pt]
            \noalign{\smallskip}
	    \hline
            \hline
            \noalign{\smallskip}
            \noalign{\smallskip}
	\multicolumn{3}{c}{model: wabs$_{\rm Gal}~\times$ 
		tbvarabs $\times$ (bb $+$ po)}\\
            \noalign{\smallskip}
            \hline
            \noalign{\smallskip}
            \noalign{\smallskip}
            	$n_{\rm H}~(10^{21}~{\rm cm}^{-2})$ & 
			$(1.6^{+0.7}_{-0.4})$ 
			& $(1.6^{+0.7}_{-0.4})~^{\mathrm{(a)}}$ \\[2pt]
		$kT_{\rm bb}~({\rm keV})$ &  $0.11^{+0.02}_{-0.01}$
			& $0.12^{+0.01}_{-0.02}$ \\[2pt]
		$K_{\rm bb}~(10^{-5})$ & $1.9^{+1.8}_{-0.9}$ 
			& $1.7^{+1.9}_{-0.8}$\\[2pt]
		$\Gamma$  &  $2.88^{+0.42}_{-0.37}$ & 
			$2.66^{+0.18}_{-0.19}$\\[2pt]
		$K_{\rm po}~(10^{-4})$ & 
			$3.0^{+0.9}_{-0.7}$ & $6.0^{+1.0}_{-0.8}$\\
            \noalign{\smallskip}
            \hline
            \noalign{\smallskip}
		$\chi^2_\nu$ & \multicolumn{2}{c}{~~~~~~~$0.72\,(130.9/182)$}  \\[2pt] 
		$f_{0.3{\rm -}12{\rm ,\,obs}}
			~(10^{-12}~{\rm erg~cm}^{-2}~{\rm s}^{-1})$
			&  $1.1^{+0.2}_{-0.3}$ & $2.0^{+0.2}_{-0.3}$\\[2pt] 
		$f_{0.3{\rm -}12}~(10^{-12}~{\rm erg~cm}^{-2}~{\rm s}^{-1})$
			&  $2.6^{+0.8}_{-0.3}$ & $4.0^{+0.9}_{-0.4}$\\[2pt]
		$f^{\mathrm{bb}}_{0.3{\rm -}12}~
			(10^{-12}~{\rm erg~cm}^{-2}~{\rm s}^{-1})$
			& $1.1^{+0.5}_{-0.2}$ & $1.0^{+0.7}_{-0.1}$  \\[2pt]
		$f^{\mathrm{po}}_{0.3{\rm -}12}~
			(10^{-12}~{\rm erg~cm}^{-2}~{\rm s}^{-1})$
			&  $1.5^{+0.3}_{-0.2}$ & $3.0^{+0.3}_{-0.3}$\\[2pt]
            	$L_{0.3{\rm -}12}~(10^{40}~{\rm erg~s}^{-1})$ & $0.7$ & $1.1$\\
            \noalign{\smallskip}
	    \hline
         \end{tabular}
\begin{list}{}{}
\item[$^{\mathrm{a}}$] assumed to be equal for the two states
%\item[$^{\mathrm{b}}$] in units of $10^{-12}~{\rm erg~cm}^{-2}~{\rm s}^{-1}$
%\item[$^{\mathrm{c}}$] in units of $10^{40}~{\rm erg~s}^{-1}$
\end{list}
\end{centering}
   \end{table}
\end{centering}

%----------------------------------------------------------- S_vib
   \begin{figure}
  \includegraphics[angle=-90, width=8.8cm]{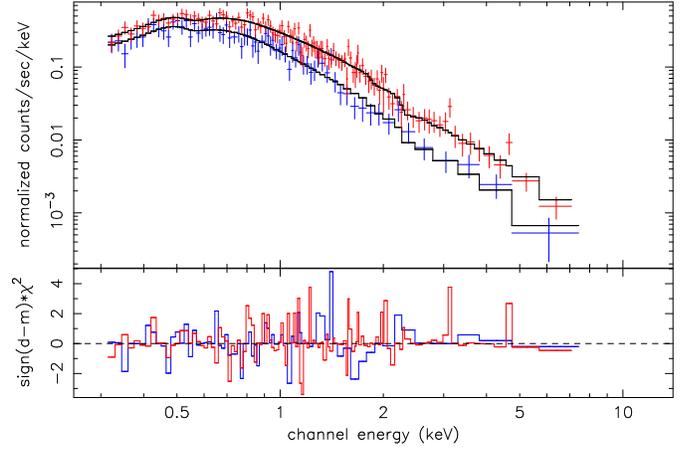}
      \caption{Coadded EPIC spectra of the ULX on 2001 Jan 28, 
separated into the ``low-flux'' and ``high-flux'' states (blue and red 
datapoints, respectively). See text for details. The spectra 
are well fitted by a {\tt {bb+po}} model: we argue that 
the blackbody component is consistent with being unchanged 
in the two states, while the variability is due to the power-law component. 
A metal abundance $Z = 0.07\,Z_{\odot}$ 
was assumed for the intrinsic absorber. See Table 8 for the fit parameters.       }
         \label{Fig1}
   \end{figure}
%
%______________________________________________________________

%----------------------------------------------------------- S_vib
   \begin{figure}
  \includegraphics[angle=-90, width=8.8cm]{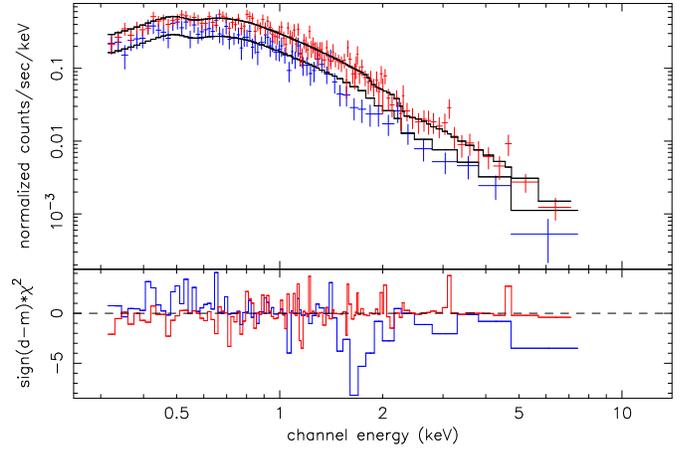}
      \caption{Same datapoints as in Fig.~6. The fitted {\tt {bb+po}}
model shows that the change in flux between low and high states 
cannot be due simply to a normalization factor: the spectrum in 
the low state is softer.}
         \label{Fig1}
   \end{figure}
%
%______________________________________________________________

\begin{centering}
   \begin{table}
      \caption[]{As in Table 6 and 7, for a {\tt{bknpo}} model.
We assumed $Z = 0.07\,Z_{\odot}$.}
         \label{table1}
\begin{centering}
         \begin{tabular}{lrr}
            \hline
            \hline
            \noalign{\smallskip}
            Parameter    & ``Low'' value & ``High'' value   \\[2pt]
            \noalign{\smallskip}
	    \hline
            \hline
            \noalign{\smallskip}
            \noalign{\smallskip}
	\multicolumn{3}{c}{model: wabs$_{\rm Gal}~\times$ 
		tbvarabs $\times$ bknpo}\\
            \noalign{\smallskip}
            \hline
            \noalign{\smallskip}
            \noalign{\smallskip}
            	$n_{\rm H}~(10^{21}~{\rm cm}^{-2})$ & 
			$(<0.6)$ 
			& $(<0.6)~^{\mathrm{(a)}}$ \\[2pt]
		$\Gamma_1$ &  $-1.1^{+2.4}_{-1.8}$
			& $-1.3^{+2.0}_{-0.2}$ \\[2pt]
		$E_{\rm br}~({\rm keV})$ & $0.55^{+0.09}_{-0.05}$ 
			& $0.56^{+0.06}_{-0.05}$\\[2pt]
		$\Gamma_2$  &  $3.25^{+0.21}_{-0.16}$ & 
			$2.79^{+0.12}_{-0.08}$\\[2pt]
		$K_{\rm bp}~(10^{-3})$ & 
			$4.9^{+10.1}_{-3.9}$ & $7.1^{+13.5}_{-5.0}$\\
            \noalign{\smallskip}
            \hline
            \noalign{\smallskip}
		$\chi^2_\nu$ & \multicolumn{2}{c}{~~~~~~~$0.75\,(137.1/182)$}  \\[2pt] 
		$f_{0.3{\rm -}12{\rm ,\,obs}}
			~(10^{-12}~{\rm erg~cm}^{-2}~{\rm s}^{-1})$
			&  $1.0^{+0.1}_{-0.1}$ & $2.0^{+0.1}_{-0.1}$\\[2pt] 
		$f_{0.3{\rm -}12}~(10^{-12}~{\rm erg~cm}^{-2}~{\rm s}^{-1})$
			&  $1.4^{+0.2}_{-0.1}$ & $2.4^{+0.4}_{-0.1}$\\[2pt]
            	$L_{0.3{\rm -}12}~(10^{40}~{\rm erg~s}^{-1})$ & $0.4$ & $0.7$\\
            \noalign{\smallskip}
	    \hline
         \end{tabular}
\begin{list}{}{}
\item[$^{\mathrm{a}}$] assumed to be equal for the two states
%\item[$^{\mathrm{b}}$] in units of $10^{-12}~{\rm erg~cm}^{-2}~{\rm s}^{-1}$
%\item[$^{\mathrm{c}}$] in units of $10^{40}~{\rm erg~s}^{-1}$
\end{list}
\end{centering}
   \end{table}
\end{centering}

\section{Conclusions}

Our spectral analysis, based on five {\it XMM-Newton}/EPIC observations 
%of the ULX in NGC\,5408 
between 2001 July and 2003 January, 
shows a thermal component 
with $kT_{\rm bb} \approx 0.12$--$0.14$ keV, significantly detected 
at all epochs, in addition to a soft power-law 
($\Gamma \approx 2.6$--$2.9$). A broken power-law model, 
as suggested in Kaaret et al.~(2003), is acceptable at some epochs, 
but is not consistent with the full series of observations, 
and is generally worse than a blackbody plus power-law model 
or a Comptonised blackbody model. 

Ruling out a simple broken power-law model 
does not rule out the possibility 
that the X-ray emission is due to inverse-Compton 
emission from a jet, enhanced by relativistic Doppler boosting 
(Kaaret et al.~2003). It was suggested in Georganopoulos (2002), 
and Georganopoulos \& Kazanas (2003), 
that a more realistic jet spectrum would have a curvature 
around the break energy, thus mimicking a soft thermal component. 
Spectral analysis over a larger energy range is required  
to test this possibility.

In any case, the X-ray spectrum is consistent 
with a more massive accreting BH in a high/soft or very high 
state. Given its inferred isotropic luminosity 
of $\approx 10^{40}$ erg s$^{-1}$ in the $0.3$--$12$ keV band,  
a mass $\ga 100 M_{\odot}$ is required to satisfy the Eddington limit.
Thus, the ULX in NGC\,5408 appears very similar to a group 
of ``canonical'' ULXs in nearby dwarf and spiral galaxies, 
%such as (***Ho II, Ho IX, NGC4559, NGC1313 ***) 
with a low-temperature thermal component at  
$kT_{\rm bb} \sim 0.1$ keV and emitted luminosity 
$\sim (1$--$2) \times 10^{40}$ erg s$^{-1}$. 
In all these systems, if the soft thermal component is interpreted 
as the emission from the inner part of a standard accretion 
disk (Shakura \& Sunyaev 1973), its temperature would imply 
masses $> 10^5 M_{\odot}$, unphysically high for a non-nuclear BH. 
However, it was also suggested (King 2003) 
that these ULXs may be accreting BHs during a super-Eddington 
accretion phase: they could have a luminosity close or even 
a factor of a few above the Eddington limit, 
accompanied by a radiatively-driven, Compton-thick outflow from 
the accretion disk. In this scenario, the soft thermal component would 
come from the photosphere of the outflow. Hence, the BH mass would 
not need to be higher than $\sim 20$--$50 M_{\odot}$, 
still consistent with a stellar origin. 

We note that this ULX is located in a very metal-poor dwarf galaxy 
($Z \approx 0.07 Z_{\odot}$). An association between ULXs 
and metal-poor environments was suggested 
by Pakull \& Mirioni (2002). Low metal abundance implies a reduced 
mass-loss rates in the radiatively-driven wind from the O-star progenitor  
($\dot{M}_{\rm w} \sim Z^{0.85}$: Vink et al.~2001; 
see also Bouret et al.~2003); this leads to a more massive stellar core, 
which may then collapse into a more massive BH, 
via normal stellar evolution.
Hence, low abundances may explain the formation 
of isolated BHs with masses up to $\approx 50 M_{\odot}$ 
(ie, up to $\approx$ half of the mass of the progenitor star). 

%The mechanisms of formation 
%and the mass of the BH and of the donor star in such systems are still 
%unclear (see, eg, ***). 
%We find no evidence of optically-thin thermal plasma emission 
%from the source, which could have been evidence of an underlying 
%supernova remnant. (although 
%some SNR have a featureless X-ray spectrum, 
%cite examples).

We found a moderate long-term spectral variability between the various 
epochs of the {\it XMM-Newton} observations, 
with a ratio between the maximum and minimum fluxes 
of $\approx 1.4$. This is consistent with the long-term 
behaviour of the source from previous {\it Einstein}, 
{\it ROSAT}, {\it ASCA} and {\it Chandra} observations (Kaaret et al.~2003).
Our preliminary timing analysis shows rapid variability and flaring-like 
behaviour, particularly in the 2003 Jan observation. 
The variability is larger in the hard band ($E > 1.5$ keV), where the flux 
is consistent with zero during the dips.

The {\tt {bb+po}} spectral model offers a clean, simple 
interpretation of the flaring behaviour. In this scenario, 
we have interpreted the lightcurves as the combination of a rapidly variable 
power-law component, plus an additional, non-variable thermal component 
below 1.5 keV, consistent with our spectral analysis. 
The power-law flux appears to be suppressed during the dips, 
when the X-ray spectrum becomes softer; 
this is supported by flux-dependent spectral analysis. 
The power-law X-ray emission in accreting BH binaries is generally 
explained as inverse-Compton scattering of soft disk photons  
in a hot corona or Compton cloud, located either above the inner 
part of the accretion disk, or as a quasi-spherical region 
inside the truncation radius of the disk. 

A flaring behaviour with soft dips or transitions 
has also been observed in Galactic microquasars 
such as GRS 1915$+$105 (Naik et al.~2001) 
and, in one case, XTE J1550$-$564 (Rodriguez et al.~2003). It was interpreted 
(Vadawale et al.~2003) as evidence that the matter responsible for the Comptonised 
component is recurrently ejected from the inner regions;
the ejected matter may be responsible for the optically-thin 
synchrotron radio emission in those systems.
%A repeated succession of soft dips and flares was also observed 
%in the Galactic BH candidate GRS 1915$+$105, and was interpreted 
%(Vadawale et al.~2001) as evidence that the matter responsible 
%for the Comptonised component is recurrently ejected 
%from the inner regions. The ejected matter would produce 
%flaring or oscillating synchrotron radio emission 
%(Vadawale et al.~2003). 
A more detailed discussion of 
this hypothesis, and of whether it may explain the steep-spectrum 
radio emission detected by the Australia Telescope Compact Array at the 
position of the ULX (Kaaret et al.~2003; Stevens et al.~2002), 
%is beyond the scope of this Letter and 
will be addressed in further work.

The power density spectrum for this ULX is 
flat below $\approx 2.5$ mHz, and has a slope 
of $\approx -1$ at higher frequencies. 
This is similar to what is found in ``canonical'' Galactic 
BH candidates and AGN: it is believed that the break frequency 
is inversely proportional to the mass of the accreting BH 
(eg, Belloni \& Hasinger 1990; Nowak et al.~1999; Uttley et al.~2002; 
Markowitz et al.~2003; Cropper et al.~2004). For example, 
the break in the power density spectrum of Cyg X-1 occurs at 
$\nu \sim 0.4$--$0.04$ Hz in different accretion states 
(Nowak et al.~1999). This suggests that the mass 
of the BH powering the ULX in NGC\,5408 is $\sim 10^2 M_{\odot}$. 
%or, more precisely, $M \approx 400 M_{\odot}$ if we adopt 
%the scaling relation of Uttley et al.~(2002). 
This is also in good agreement with the inferred luminosity, 
if we require that it does not exceed the Eddington limit.
Such a high mass would rule out (cf.~Fig.~3 of Kaaret et al.~2004) 
an association of this ULX with the young star clusters 
in the starburst region of NGC\,5408, located $\approx 300$ pc 
to the north-west.
However, we are aware that the relation between 
break frequency and BH mass is very uncertain, 
and the power-density spectra are often 
more complicated than a simple broken power law 
(see, eg, the case of GRS 1915$+$105: Morgan et al.~1997).

In conclusion, we argue that both the spectral and timing results 
from our {\it XMM-Newton} study are consistent with the possibility 
that this ULX is at the upper end of the mass range for BHs 
of stellar origin, a few times more massive than ``standard'' BH candidates 
such as Cyg X-1. In that case, its luminosity would be 
very close or, more likely, slightly higher than its Eddington limit, 
and a strong outflow would be expected from its accretion disk.
Doppler boosting in a relativistic jet is not required 
to explain the observed properties, although 
we cannot rule it out.

%I think it is interesting that the NGC5408 ULX is also a little bit away
%(~300pc) from the starburst region (as discussed in the case of X-ray 
%sources in other starburst galaxies by Kaaret et al 2004, MNRAS 348, L28),
%but comparing with Fig 3 of that paper, our source clearly violates the
%luminosity/separation trend seen. 

\begin{acknowledgements}

We thank Manfred Pakull for preparing the observations 
and for discussions, and Kinwah Wu for comments on the revised version.
We also thank the referee (Phil Kaaret) for his constructive 
criticism of the first version.

\end{acknowledgements}


\begin{thebibliography}{}
   \bibitem[]{} 

   \bibitem[1996]{arnaud} Arnaud, K. A. 1996, Astronomical Data Analysis 
	Software and Systems V, 
%eds G. Jacoby \& J. Barnes, 
	ASP Conf. Series, 101, 17

   \bibitem[2001]{belloni} Belloni, T. 2001, preprint (astro-ph/0112217)

   \bibitem[1990]{belloni} Belloni, T., \& Hasinger, G. 1990, A\&A, 227, L33

   \bibitem[1995]{blackburn} Blackburn, J. K. 1995, Astronomical Data 
	Analysis Software and Systems IV, ASP Conference Series, 77, 367

%   \bibitem[1972]{bohuski} Bohuski, T. J., Burbidge, E. M., Burbidge, G. R., 
%	\& Smith, M. G. 1972, ApJ, 175, 329

   \bibitem[1999]{bouret} Bouret, J.-C., Lanz, T., Hillier, D. J., et al.~2003, 
	ApJ, 595, 1182

   \bibitem[1999]{colbert} Colbert, E. J. M., \& Mushotsky, R. F. 1999, 
	ApJ, 519, 89

   \bibitem[2004]{cropper} Cropper, M. S., Soria, R., Mushotzky, R. F., 
	et al.~2004, MNRAS, 349, 39

   \bibitem[2004]{dewangan} Dewangan, G. C., Miyaji, T., Griffiths, R. E., 
	\& Lehmann, L. 2004, ApJ, submitted (astro-ph/0401223)

   \bibitem[1990]{dickey} Dickey, J. M., \& Lockman, F. J. 1990, 
	ARA\&A, 28, 215

   \bibitem[1995]{fabbiano} Fabbiano, G. 1992, ARA\&A, 27, 87

   \bibitem[1993]{fabian} Fabian, A. C., \& Ward, M. J. 1993, MNRAS, 263, L51

   \bibitem[2002]{george} Georganopoulos, M., Aharonian, F. A., 
	\& Kirk, J. G.~2002, A\&A, 388, L25

   \bibitem[2003]{george2} Georganopoulos, M., \& Kazanas, D. 2003, ApJ, 589, L5

   \bibitem[1970]{hamilton} Hamilton, W. C. 1965, Acta Cryst., 18, 502

   \bibitem[1970]{himmel} Himmelblau, D. M. 1970, Process Analysis by Statistical 
	Methods, New York: John Wiley, pp. 214--222

   \bibitem[2003]{kaaret1} Kaaret, P., Corbel, S., Prestwich, A. H., 
	\& Zezas, A. 2003, Science, 299, 365

   \bibitem[2004]{kaaret2} Kaaret, P., et al.~2004, MNRAS, 348, L2

   \bibitem[2002]{karachentsev} Karachentsev, I. D., Sharina, M. E., 
	Dolphin, A. E., et al.~2002, A\&A, 385, 21

   \bibitem[2002]{king1} King, A. R. 2002, MNRAS, 335, L13

   \bibitem[2003]{king2} King, A. R. 2003, in the Proc. of the 2nd {\it BeppoSAX} 
	Meeting: "The Restless High-Energy Universe" (Amsterdam, May 5-8, 2003), 
	E. P. J. van den Heuvel, J. J. M. in 't Zand, and R. A. M. J. Wijers Eds, 
	astro-ph/0309524

   \bibitem[2001]{kording} K\"{o}rding, E., Falcke, H., \& Markoff, S. 2002, 
	A\&A, 382, L13

   \bibitem[2001]{madau} Madau, P., \& Rees, M. J. 2001, ApJ, 551, L27

   \bibitem[2003]{markowitz} Markowitz, A., Edelson, R., Vaughan, S., 
	et al.~2003, ApJ, 593, 96

   \bibitem[2003]{miller} Miller, J. M., Fabbiano, G., Miller, M. C., 
	\& Fabian, A. C. 2003, ApJ, 585, L37

   \bibitem[2001]{morgan} Morgan, E. H., Remillard, R. A., \& Greiner, J. 1997, 
	ApJ, 482, 993

   \bibitem[2001]{naik} Naik, S., Agrawal, P. C., Rao, A. R., et al. 2001, ApJ, 546, 1075

   \bibitem[1999]{nowak} Nowak, M. A., Vaughan, B. A., Wilms, J., 
	Dove, J. B., \& Begelman, M. C. 1999, ApJ, 510, 874

   \bibitem[2003]{page} Page, M. J., Davis, S. W., \& Salvi, N. J. 2003, 
	MNRAS, 343, 1241

   \bibitem[2002]{pakull}
	Pakull, M. W., \& Mirioni, L. 2002, to appear 
	in the proceedings of the symposium 'New Visions 
	of the X-ray Universe', 26-30 November 2001, 
	ESTEC, The Netherlands (astro-ph/0202488)

   \bibitem[1982]{prince} Prince, E. 1982, Acta Cryst., B38, 1099

   \bibitem[2004]{zwart} Portegies Zwart, S. F., Baumgardt, H., Hut, P., 
	Makino, J., \& McMillan, S. L. W. 2004, Nature, in press (astro-ph/0402622)

   \bibitem[2000]{roberts} Roberts, T. P., \& Warwick, R. S. 2000, 
	MNRAS, 315, 98

   \bibitem[2003]{rodriguez} Rodriguez, J., Corbel, S., 
	\& Tomsick, J. A. 2003, ApJ, 595, 1032

   \bibitem[1973]{shakura} Shakura, N. I., \& Sunyaev, R. A. 1973, A\&A, 24, 337

   \bibitem[1999]{shrader} Shrader, C. R., \& Titarchuk, L. G. 
	1999, ApJ, 521, L21

   \bibitem[2002]{stevens} Stevens, I. R., Forbes, D. A., 
	\& Norris, R. P. 2002, MNRAS, 335, 1079

   \bibitem[1982]{stewart} Stewart, G. C., Fabian, A. C., Terlevich, R. J., 
	\& Hazard, C. 1982, MNRAS, 200, 61P

   \bibitem[2003]{uttley} Uttley, P., McHardy, I. M., 
	\& Papadakis, I. E.~2002, MNRAS, 332, 231

   \bibitem[2003]{vadawale1} Vadawale, S. V., Rao, A. R., 
	Naik, S., et al.~2003, ApJ, 597, 1023

   \bibitem[2003]{vink} Vink, J. S., de Koter, A., \& Lamers, H. J. G. L. M. 
	 2001, A\&A, 369, 574

   \bibitem[1953]{williams} Williams, E. J., \& Kloot, N. H. 1953, Aust. J. Appl. Sci., 4, 1

   \bibitem[2000]{wilms} Wilms, J., Allen, A., \& McCray, R. 2000, ApJ, 542, 914

%   \bibitem[2001]{vadawale2} Vadawale, S. V., Rao, A. R., Nandi, A., 
%	\& Chakrabarti, S. K.~2001, A\&A, 370, L17

\end{thebibliography}
\end{document}